\documentclass[sigchi]{acmart}

\usepackage{ctable}
\usepackage{pifont}

\usepackage{multirow}
\usepackage{balance}
\usepackage{booktabs}
\usepackage{hyphsubst}
\usepackage{graphicx}
\usepackage{amsmath}
\usepackage{url}
\usepackage{hhline}
\usepackage[caption = false]{subfig}

\newcommand{\para}[1]{\vspace{2mm}\noindent\textbf{#1}}
\newcommand{\subpara}[1]{\textit{\textbf{#1}}}

\DeclareMathOperator*{\argmax}{arg\,max}
\DeclareMathOperator*{\argmin}{arg\,min}

\setcopyright{rightsretained}
\copyrightyear{2020}
\acmYear{2020}
\acmDOI{https://doi.org/10.48550/arXiv.2003.10699}
\acmISBN{}

\acmConference[IUI '20 Workshops]{IUI '20 Workshops}{March 17, 2020}{Cagliari, Italy}
\acmBooktitle{IUI '20 Workshops, March 17, 2020, Cagliari, Italy}

\begin{document}

\title[Utilizing Human Memory Processes to Model Genre Preferences]{Utilizing Human Memory Processes to Model Genre Preferences for Personalized Music Recommendations}

\author{Dominik Kowald}
\authornote{Both authors contributed equally to this work.}
\affiliation{%
  \institution{Know-Center GmbH}
  \city{Graz, Austria} 
}
\email{dkowald@know-center.at}

\author{Elisabeth Lex}
\authornotemark[1]
\affiliation{%
  \institution{Graz University of Technology}
  \city{Graz, Austria} 
}
\email{elisabeth.lex@tugraz.at}

\author{Markus Schedl}
\affiliation{%
  \institution{Johannes Kepler University Linz}
  \city{Linz, Austria} 
}
\email{markus.schedl@jku.at}

\begin{abstract}
In this paper, we introduce a psychology-inspired approach to model and predict the music genre preferences of different groups of users by utilizing human memory processes. These processes describe how humans access information units in their memory by considering the factors of (i) past usage frequency, (ii) past usage recency, and (iii) the current context. Using a publicly available dataset of more than a billion music listening records shared on the music streaming platform Last.fm, we find that our approach provides significantly better prediction accuracy results than various baseline algorithms for all evaluated user groups, i.e., (i) low-mainstream music listeners, (ii) medium-mainstream music listeners, and (iii) high-mainstream music listeners. Furthermore, our approach is based on a simple psychological model, which contributes to the transparency and explainability of the calculated predictions. 
\end{abstract}

\maketitle


\section{Introduction}
\label{s:intro}
Computational models of user preferences are crucial elements of music recommender systems~\cite{schedl2015music} to tailor recommendations to the preferences of the user. Such user models are typically derived from the listening behavior of the users, i.e., their interactions with music artifacts, content features of music~\cite{zangerlecontent}, or hybrid combinations of both. Research in music psychology~\cite{north2008social} has shown that a wide range of factors impact music preferences~\cite{schedl2015music}, such as emotional state~\cite{cantor1973effect,juslin2001music}, a user's current context~\cite{rentfrow2003re}, or a user's personality~\cite{rentfrow2003re,schedl_etal:taffc:2017}. Several aspects make the modeling of music preferences challenging, such as, e.g., that music consumption is context-dependent and serves various purposes for listeners~\cite{Schedl2018challengesmrs}. Also, recent research~\cite{kowald2020unfairness} has verified that classic music recommendation approaches suffer from popularity bias, i.e., they are biased to the mainstream that is prevalent in a music community. As a result, listeners of non-mainstream music receive less relevant recommendations compared to listeners of popular, mainstream music~\cite{bauer2019global,schedl_bauer:jmm:2018,schedl2017distance,Oord2013:DCM}. 

\begin{table*}[t]
  \setlength{\tabcolsep}{4.0pt}    
  \centering
    \begin{tabular}{r|ccccc|ccccc}
    \specialrule{.2em}{.1em}{.1em}
        User Group                 & $|U|$            & $|A|$     & $|G|$        & $|LE|$        & $|GA|$            & $|GA|/|LE|$ & $|G|/|U|$ & $Avg. MS$ & $Avg. Age$ & $M/F$       \\\hline 
        LowMS  &    1,000            &    82,417    &    931        &    6,915,352    &    14,573,028    &    2.107  & 85.771 & .125 & 24.582 & 74\%/26\%        \\\hline
        MedMS     & 1,000          & 86,249                 & 933        & 7,900,726 & 20,264,870    & 2.565          & 126.439 & .379 & 25.352 & 68\%/32\%       \\\hline
        HighMS    &    1,000    &    92,690    &    973        &    8,251,022    &    22,498,370    &    2.727          & 186.010 & .688 & 21.486 & 65\%/35\%            \\
        \specialrule{.2em}{.1em}{.1em}                                
    \end{tabular}
    \caption{Dataset statistics for the LowMS, MedMS, and HighMS Last.fm user groups. Here, $|U|$ is the number of distinct users, $|A|$ is the number of distinct artists, $|G|$ is the number of distinct genres, $|LE|$ is the number of listening events, $|GA|$ is the number of genre assignments, $|GA|/|LE|$ is the average number of genre assignments per LE,  $|G|/|U|$ is the average number of genres a user has listened to, $Avg. MS$ is the average mainstreaminess value, $Avg. Age$ is the average age of users in the group and $ M/F$ is the users' male/female ratio.}
  \label{tab:datasets}
\end{table*}

In this paper, we introduce a psychology-inspired approach to model and predict the music genre preferences of users. We base our approach on research in music psychology that found music liking being positively influenced by prior exposure to the  music~\cite{pereira2011music,schubert2007influence}. This has been attributed to the \emph{mere exposure effect} or \emph{familiarity principle}~\cite{zajonc1968attitudinal}, i.e., users tend to establish positive preferences for items to which they are frequently and consistently exposed. Our idea is to computationally model prior exposure to music genres using the activation equation of human memory from the cognitive architecture \emph{Adaptive Control of Thought--Rational} (ACT-R)~\cite{anderson1991reflections,anderson2004integrated}. The activation equation determines the usefulness of a memory unit (i.e., its \emph{activation}) for a user in the current context, based on how frequently and recently a user accessed it in the past as well as how important this unit is in the current context. In our previous work, we have employed a specific part of the activation equation, namely the Base-Level-Learning (BLL) equation, to recommend music artists~\cite{ismir_lfm_2019}. The BLL equation computes the base-level activation of a memory unit based on how frequently and recently a user has accessed it in the past, following a time-dependent decay in the form of a power-law distribution. A high base-level activation means that the memory unit is vital for the user and, thus, can be more easily retrieved from her memory. However, in this work~\cite{ismir_lfm_2019}, we did not implement the full activation equation as we left out the associative activation that tunes the base-level activation of the memory unit to the current context. 

In the present paper, we extend our previous model and utilize the associative activation for music genre predictions. This helps us tune the predictions to the current context of the user. As the current context, we utilize the set of genres that are assigned to the most recently listened artist of a user. On a publicly available dataset of Last.fm music listening histories, we model the genre preferences of users from three different groups, which we extract using behavioral data in the form of music listening events: (i) LowMS, i.e., listeners of niche music (low mainstreaminess), (ii) HighMS, i.e., listeners of mainstream music (high mainstreaminess), and (iii) MedMS, i.e., listeners of music that lies in-between (medium mainstreaminess). We introduce the $ACT_{u,a}$ approach that employs the full activation equation to take into account the current context of the user, which we define as the user's current genre preference. We compare the efficacy of $ACT_{u,a}$ to a variant, i.e., $BLL_u$, that uses only the BLL equation to model the past usage frequency (i.e., popularity) and recency (i.e., time). Furthermore, we compare both approaches to five baselines, including two collaborative filtering variants, mainstream-aware genre modeling, popularity-aware genre modeling, as well as time-based genre modeling.

The contributions of our work are two-fold. Firstly, we propose $ACT_{u,a}$, as an extension to $BLL_u$, to model and predict the genre preferences of users. Secondly, we evaluate the efficacy of both $BLL_u$ and $ACT_{u,a}$ on three different groups of Last.fm users, which we separate based on the distance of their listening behavior to the mainstream: (i) LowMS, (ii) MedMS, and (iii) HighMS. We find that both $BLL_u$ and $ACT_{u,a}$ outperform the five baseline methods in all three groups, with $ACT_{u,a}$ achieving the significantly highest performance. Our results also show that with both $BLL_u$ and $ACT_{u,a}$, we can specifically improve the prediction performance for the users in the LowMS group. In other words, we can serve better the music consumers, whose prediction quality suffers the most from popularity bias. Also, both $BLL_u$ and $ACT_{u,a}$ are based on a psychological theory, whose computational model is transparent and explainable and not a black box. 

\section{Data and Approach}
\label{s:method}
In this section, we describe the Last.fm dataset as well as our music genre modeling and prediction approaches.

\subsection{Dataset}
In this paper, we use the publicly available \emph{LFM-1b} dataset\footnote{\url{http://www.cp.jku.at/datasets/LFM-1b/}} of music listening information shared by users of the online music platform Last.fm. \emph{LFM-1b} contains listening histories of more than 120,000 users, which sums up to over 1.1 billion listening events (LEs) collected between January 2005 and August 2014. Each LE contains a user identifier, the artist, the album, the track name, and a timestamp~\cite{schedl2016lfm}.
Furthermore, the \emph{LFM-1b} dataset contains demographic data of the users such as country, age, gender, and a mainstreaminess score, which is defined as the overlap between a user's personal listening history and the aggregated listening history of all Last.fm users in the dataset. Thus, the mainstreaminess score reflects a user's inclination to music listened to by the Last.fm mainstream listeners (i.e., the ``average'' Last.fm listener)~\cite{schedl2015tailoring}.

\para{User groups.}
In order to study different types of users, we use this mainstreaminess score to split the \emph{LFM-1b} dataset into three equally sized user groups based on their mainstreaminess (i.e., low, medium, and high). 
Specifically, we sort all users based on their mainstreaminess score and assign the 1,000 users with the lowest scores to the low-mainstream group (i.e., \textit{LowMS}), the 1,000 users with scores around the median mainstreaminess (= .379) to the medium-mainstream group (i.e., \textit{MedMS}), and the 1,000 users with the highest scores to the high-mainstream group (i.e., \textit{HighMS}).

In our study, we consider only users with a minimum of 6,000 and a maximum of 12,000 LEs. We choose these thresholds based on the average number of LEs per user in the dataset, which is 9,043, as well as the kernel density distribution of the data. With this method, on the one hand, we exclude users with too little data available for training our algorithms (i.e., users with less than 6,000 LEs), and on the other hand, we exclude so-called power listeners (i.e., users with more than 12,000 LEs) that might distort our results. 
Table~\ref{tab:datasets} summarizes the statistics and characteristics of our three user groups. We see that, even if we only consider 1,000 users per group, we have a sufficient amount of LEs, i.e., between 6.9 to 8.3 million, to train and test our music genre modeling and prediction approaches. Further characteristics of our user groups are as follows:

\subpara{(i) LowMS.} 
The LowMS group represents the $|U|$~=~1,000 users with the smallest mainstreaminess scores. These users have an average mainstreaminess value of $Avg. MS$~=~.125. LowMS contains $|A|$~=~82,417 distinct artists, $|LE|$~=~6,915,352 listening events, $|G|$~=~931 genres, and $|GA|$~=~14,573,028 genre assignments. Interestingly, the male/female ratio is the least evenly distributed one in this group (i.e., $M/F$~=~74\%/26\%).

\subpara{(ii) MedMS.}
The MedMS group consists of the $|U|$~=~1,000 users with mainstreaminess scores around the median and thus, lying between the ones of the LowMS and HighMS groups. This group has an average mainstreaminess value of $Avg. MS$~=~.379. 
The majority of dataset statistics of this group lies between the ones of the LowMS and HighMS users, except for the average age, which is the highest for the MedMS users (i.e., $Avg. Age$~=~25.352 years).

\subpara{(iii) HighMS.}
The HighMS group represents the $|U|$~=~1,000 users in the \emph{LFM-1b} dataset with the highest mainstreaminess scores ($Avg. MS$~=~.688). These users are not only the youngest ones (i.e., $Avg. Age$~=~21.486 years) but also listen to the highest number of distinct genres on average (i.e., $|G|/|U|$~=~186.010), indicating that music which is considered mainstream is quite diverse on Last.fm. 
Also, this user group exhibits the largest number of female listeners (i.e., $M/F$~=~65\%/35\%) and the highest number of distinct genres ($|G|$~=~973).

Additionally, we investigate the most frequent countries of the users. Here, for all three groups, the United States (US) is the dominating country. The share of US users increases with the mainstreaminess, i.e., while this share is only 14\% for LowMS and 18\% for MedMS, it is already 22\% for HighMS. Interestingly, Russia (RU, 13\%), Poland (PL, 9\%), and Japan (JP, 8\%) are frequent in the LowMS group, while the United Kingdom (UK) contributes a substantial share in the other two groups (9\% for MedMS and 14\% for HighMS). Germany (DE) is among the most popular countries in all three groups (10\% for LowMS and HighMS, 8\% for MedMS); Brazil (BR) can only be found among the most popular countries in the MedMS group (8\%); and the Netherlands (NL, 5\%) as well as Spain (ES, 4\%) can only be found in the HighMS group.

\para{Genre mapping.} 
For mapping music genres to artists, we use an extension of the \emph{LFM-1b} dataset, namely the \emph{LFM-1b UGP} dataset~\cite{schedl2017large}, which describes the genres of an artist by leveraging social tags assigned by Last.fm users. Specifically, \emph{LFM-1b UGP} contains a weighted mapping of 1,998 music genres available in the online database Freebase\footnote{\url{https://developers.google.com/freebase/} (no longer maintained)} to Last.fm artists. This database includes a fine-grained representation of musical styles, including genres such as ``Progressive Psytrance'' or ``Pagan Black Metal''.

The genre weightings for any given artist correspond to the relative frequency of tags assigned to that artist in Last.fm. For example, for the artist ``Metallica'', the top tags and their corresponding relative frequencies are ``thrash metal'' (1.0), ``metal'' (.91), ``heavy metal'' (.74), ``hard rock'' (.41), ``rock'' (.34), and ``seen live'' (.3). From this list, we remove all tags that are not part of the 1,998 Freebase genres (i.e., ``seen live'' in our example) as well as all tags with a relative frequency smaller than .5 (i.e., ``hard rock'' and ``rock'' in our example). Thus, for ``Metallica'', we end up with three genres, i.e., ``thrash metal'', ``metal'' and ``heavy metal''.

\subsection{Approach}
In this section, we describe our music genre modeling and prediction approach based on the declarative memory module of ACT-R.

\subsubsection{The Cognitive Architecture ACT-R}
ACT-R, which is short for ``Adaptive Control of Thought -- Rational'', is a cognitive architecture developed by John Robert Anderson \cite{anderson2004integrated}. ACT-R defines and formalizes the basic cognitive operations of the human mind (e.g., access to information in human memory).

\begin{figure}[t] 
   \centering
      \includegraphics[width=0.49\textwidth]{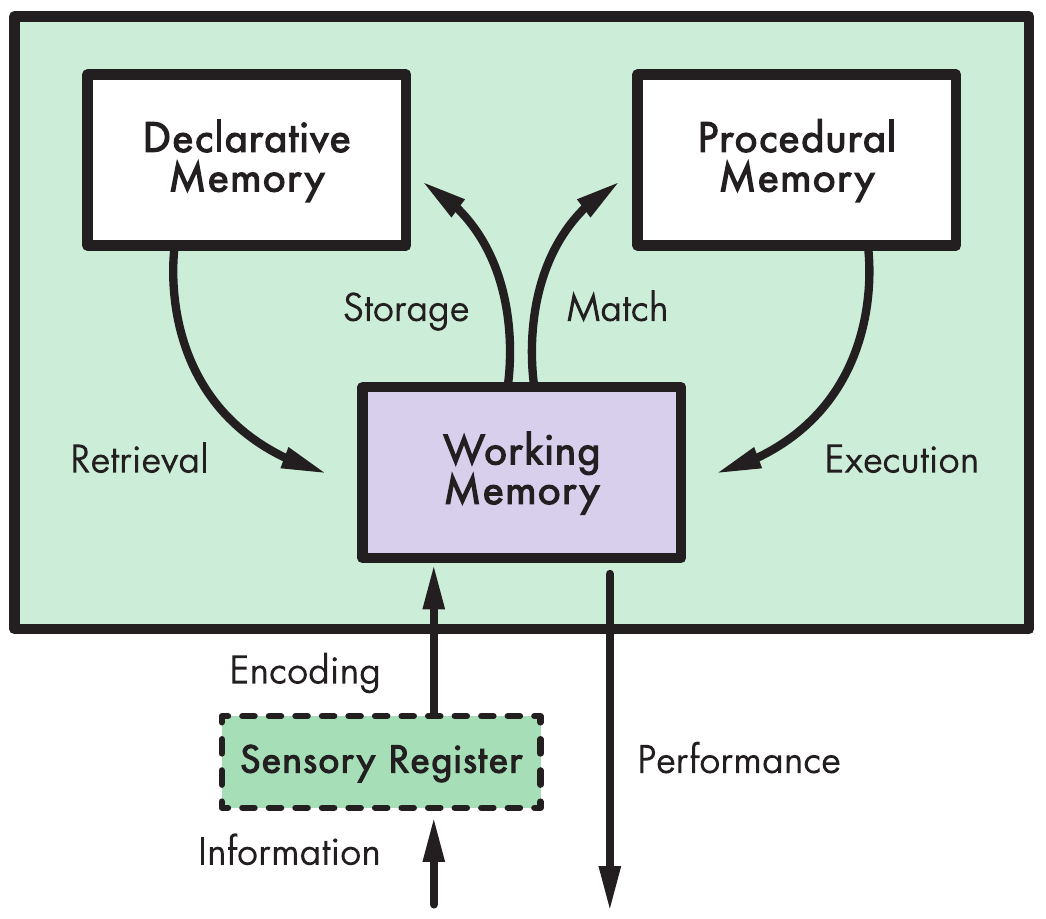}
   \caption{Schematic illustration of ACT-R. In our work, we focus on the activation equation of the declarative memory module.}
     \label{fig:actr}
\end{figure}

Figure~\ref{fig:actr} schematically illustrates the main architecture of ACT-R. In general, ACT-R differs between short-term memory modules, such as the working memory module, and long-term memory modules, such as the declarative and procedural memory modules. Using a sensory register (i.e., the ultra-short-term memory), the encoded information is passed to the short-term working memory module, which interacts with the long-term memory modules. In the case of the declarative memory, the encoded information can be stored, and already stored information can be retrieved. In the case of the procedural memory, the information can be matched against stored rules that can lead to actions~\cite{actrImage}.

\begin{figure*}[t]
   \centering
     \captionsetup[subfigure]{justification=centering}
     \subfloat[][User group: LowMS\\Linear regression: $\alpha$~=~-1.480]{ 
      \includegraphics[width=0.32\textwidth]{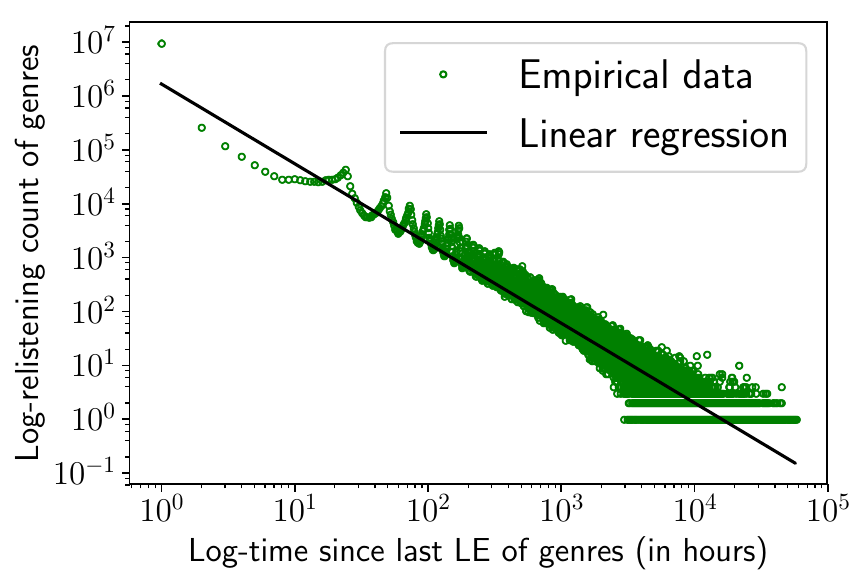} 
   }
     \subfloat[][User group: MedMS\\Linear regression: $\alpha$~=~-1.574]{ 
      \includegraphics[width=0.32\textwidth]{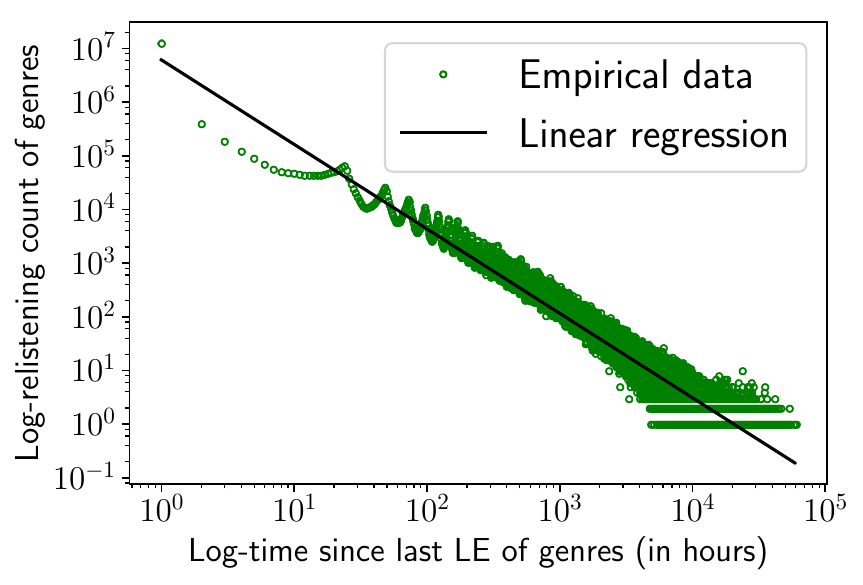} 
   }
     \subfloat[][User group: HighMS\\Linear regression: $\alpha$~=~-1.587]{ 
      \includegraphics[width=0.32\textwidth]{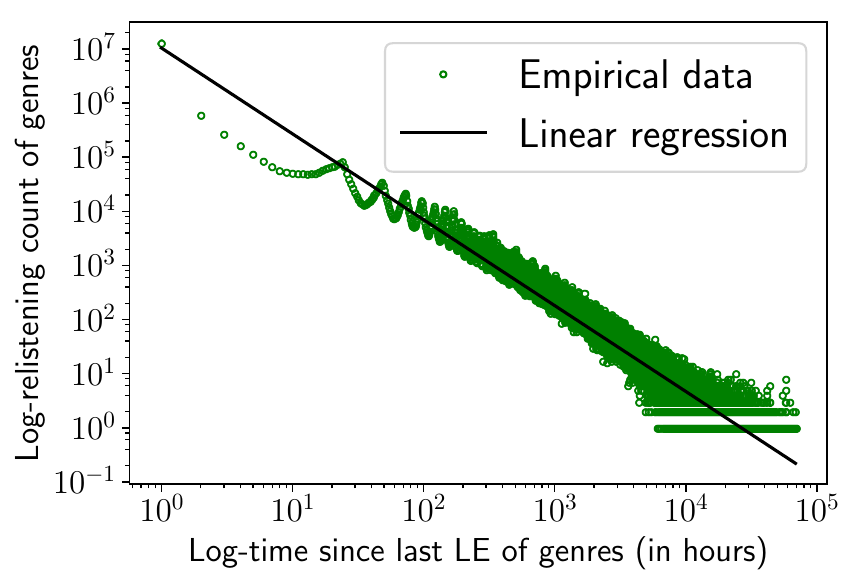} 
   }
   \caption{Calculation of the BLL equation's $d$ parameter. On a log-log scale, we plot the relistening count of the genres over the time since their last LEs. We set $d$ to the slopes $\alpha$ of the corresponding linear regression lines.}
     \label{fig:time_analysis}
\end{figure*}

Thus, declarative memory holds factual knowledge (e.g., what something is), and procedural memory consists of sequences of actions (e.g., how to do something). In our work, we focus on the declarative part, which contains the activation equation of human memory. The activation equation determines the usefulness, i.e., the activation level $A_i$, of a memory unit $i$ (e.g., a music genre in our case) for a user $u$ in the current context. It is given by:
 \begin{equation} \label{eq:A}
		A_i~=~B_i + \sum_j{W_j \cdot S_{j,i}}
  \end{equation}
Here, the $B_i$ component represents the \textit{base-level} activation and quantifies the general usefulness of the unit $i$ by considering how frequently and recently it has been used in the past. It is given by the base-level learning (BLL) equation: 
\begin{equation} \label{eq:bll}
    B_i~=~ln\left(\sum\limits_{j~=~1}\limits^{n}{t_{j}^{-d}}\right)
  \end{equation}
where $n$ is the frequency of $i$'s occurrences and $t_j$ is the time since the $j^{th}$ occurrence of $i$. The exponent $d$ accounts for the power-law of forgetting, which means that each unit's activation level caused by the $j^{th}$ occurrence decreases in time according to a power function \cite{anderson2004integrated}.

The second component of Equation \ref{eq:A} represents the \emph{associative activation} that tunes the base-level activation of the unit $i$ to the current context. The context is given by any contextual element $j$ that is relevant for the current situation. In the case of a music recommender system, that could be a music genre that the user prefers in the current situation. Through learned associations, the contextual elements are connected with $i$ and can increase $i$'s activation depending on the weight $W_j$ and the strength of association $S_{j,i}$.

\subsubsection{Modeling and Predicting Music Genre Preferences}
For modeling and predicting music genre preferences, we investigate two approaches: (i) $BLL_{u}$ based on the BLL equation to model the past usage frequency (i.e., popularity) and recency (i.e., time), and (ii) $ACT_{u,a}$ based on the full activation equation to also take the current context into account.

We start with $BLL_u$ and thus, with defining the base-level activation $B(g,u)$ for genre $g$ and user $u$ by utilizing the previously defined BLL equation:
\begin{equation} \label{eq:bllu}
    B(g,u)~=~ln\left(\sum\limits_{j~=~1}\limits^{n}{t_{u,g,j}^{-d}}\right)
\end{equation}
Here, $g$ is a genre user $u$ has listened to in the past, and $n$ is the number of times $u$ has listened to $g$. Further, $t_{u,g,j}$ is the time in seconds since the $j$\textsuperscript{th} LE of $g$ by $u$, and $d$ is the power-law decay factor, which we identify using a similar method as used in~\cite{www_hashtag_2017}. Thus, in Figure~\ref{fig:time_analysis}, for all LEs and genres in our dataset, we plot the relistening count of a genre $g$ over the time since the last LE of $g$. Then, we set $d$ to the slope $\alpha$ of the linear regression lines of this data, which leads to 1.480 for LowMS, 1.574 for MedMS, and 1.587 for HighMS.

The resulting base-level activation values $B(g, u)$ are then normalized using a simple softmax function in order to map them onto a range of [0, 1] that sums up to 1~\cite{www_hashtag_2017,Kowald2016bllhypertext}:
\begin{equation} \label{eq:sm}
    B'(g,u)~=~\frac{\exp(B(g,u))}{\sum\limits_{g' \in G_{u}}{\exp(B(g',u))}}
\end{equation}
Here, $G_u$ is the set of distinct genres listened to by $u$. Finally, $BLL_u$ predicts the top-$k$ genres $\widetilde{G_u^k}$ with highest $B'(g,u)$ values to $u$:
\begin{equation} \label{eq:rank}
    \underbrace{\widetilde{G_u^k}~=~\argmax_{g \in G_{u}}^{k}(B'(u,g))}_{BLL_u}
\end{equation}

To investigate not only the factors of frequency and time but also the current context by means of an associative activation, we implement the full activation equation (see Equation \ref{eq:A}) in the form of:
\begin{equation} \label{eq:activation}
 A(g,u,a) = B'(g,u) + \sum\limits_{c \in G_{a}}{W_c \cdot S_{c,g}}
\end{equation}
where the first part represents the base-level activation by means of the BLL equation and the second part represents the associative activation.

\begin{figure}[t] 
   \centering
      \includegraphics[width=0.49\textwidth]{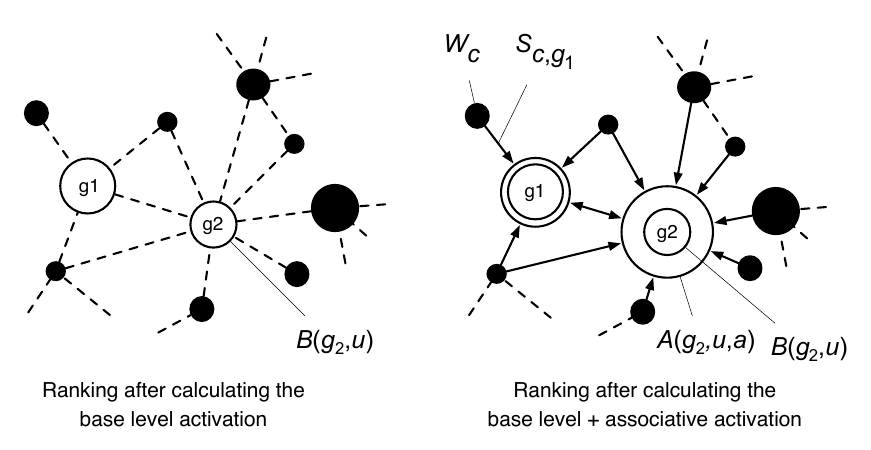}
   \caption{Example illustrating the difference between $BLL_u$ (left panel) and $ACT_{u,a}$ (right panel) based on~\cite{trattner2016modeling}. Here, unfilled nodes represent target genres $g_1$ and $g_2$, and black nodes represent genres of the last artist listened to by the target user (i.e., contextual genres). For $g_1$ and $g_2$, the node sizes represent the activation levels and for the contextual genres, the node sizes represent the attentional weights $W_c$. The association strength $S_{c,g}$ is represented by the edge lengths. While $BLL_u$ determines a higher activation level for $g_1$ than for $g_2$, $ACT_{u,a}$ gives a higher activation level to $g_2$ than to $g_1$ by also considering the associative association based on the current context.}
     \label{fig:example}
\end{figure}

To calculate the associative activation and thus, to model a user's current context, we incorporate the set of genres $G_{a}$ assigned to the most recently listened to artist $a$ by user $u$. When applying this equation in the context of recommender systems, related literature~\cite{van2009recommender} suggests using a measure of normalized co-occurrence to represent the strength of an association $S_{c,g}$. Accordingly, we define the co-occurrence between two genres as the number of artists to which both genres are assigned. We normalize this co-occurrence value according to the Jaccard coefficient:   
\begin{equation} \label{eq:cooccurrence}
	S_{c,g}~=~\frac{|A_c \cap A_g|}{|A_c \cup A_g|}
\end{equation}
where $A_c$ is the set of artists to which context-genre $c$ is assigned, and $A_g$ is the set of artists to which genre $g$ is assigned. Thus, we set the number of times two genres co-occur into relation with the number of times in which at least one of the two genres appears. In this work, we set the attentional weight $W_c$ of context-genre $c$ to 1. By doing so, we give equal weights to all genres assigned to an artist, which avoids down-ranking of less popular, but perhaps more specific, and hence more valuable, genres.

Finally, we normalize the $A(g,u,a)$ values using the aforementioned softmax function and predict the top-$k$ genres $\widetilde{G_{u}^k}$ with highest $A'(g,u,a)$ values for a given user $u$ and the genres of the user's most recently listened artist $a$ (i.e., the current context):
\begin{equation}
    \underbrace{\widetilde{G_{u}^k}~=~\argmax_{g \in G_{u}}^{k}(A'(u,g,a))}_{ACT_{u,a}}
\end{equation}

We further illustrate the difference between $BLL_u$ and $ACT_{u,a}$ in the example of Figure~\ref{fig:example}~\cite{trattner2016modeling} by showing the additional impact of the associative activation defined by the second component of the activation equation. As defined, this associative activation is evoked by the current context (i.e., the genres of the last artist the target user has listened to).

The left panel of Figure~\ref{fig:example} shows two genres, $g_1$ and $g_2$, with different base-level activation levels (illustrated by the circle size). Thus, according to $BLL_u$, $g_1$ reaches a higher base-level activation, which means a better rank, than $g_2$. This relationship changes in the right panel of Figure~\ref{fig:example}, where we consider the influence of the genres in the current context (illustrated by the black nodes). Specifically, depending on the weights $W_c$ (represented by the size of the black nodes) and strength of association $S_{c,g}$ (represented by the length of the edges), the genres in the current context spread additional associative activation to the genres $g_1$ and $g_2$. Now, according to $ACT_{u,a}$, $g_2$ receives stronger associative activation than $g_1$, which also leads to a better rank.

\section{Experiments and Results}
\label{s:results}
In this section, we describe our experimental setup, i.e., the baseline algorithms, the evaluation protocol and metrics, as well as the results of our experiments.

\subsection{Baseline Algorithms}
We compare the $BLL_u$ and $ACT_{u,a}$ approaches to five baseline algorithms: 

\para{Mainstream-based baseline: \boldmath{$TOP$}.} 
The $TOP$ approach models a user $u$'s music genre preferences using the overall top-$k$ genres of all users (i.e., the mainstream) in $u$'s user group (i.e., LowMS, MedMS, HighMS). This is given by:
\begin{equation} \label{eq:top}
    \widetilde{G_u^k}~=~\argmax_{g \in G}^{k}(|GA_{g}|)
\end{equation}
Here $\widetilde{G_u^k}$ denotes the set of $k$ predicted genres, $G$ the set of all genres, and $|GA_{g}|$ corresponds to the number of times $g$ occurs in all genre assignments $GA$ of $u$'s user group.

\para{User-based collaborative filtering baseline: \boldmath{$CF_u$}.}
User-based collaborative filtering-based approaches aim to find similar users for target user $u$ (i.e., the set of neighbors $N_u$) and predict the genres these similar users have listened to  in the past~\cite{shi_etal:compsurv:2014}. $CF_u$ is given by:
\begin{equation} \label{eq:cf}
    \widetilde{G_u^k}~=~\argmax_{g \in G(N_u)}^{k}\left(\sum\limits_{v \in N_{u}}{sim(G_{u}, G_{v}) \cdot |GA_{g, v}|}\right)
\end{equation}
where $\widetilde{G_u^k}$ denotes the set of $k$ predicted genres for user~$u$, $G(N_u)$ are the genres listened to by the set of neighbors $N_u$,\footnote{We set the neighborhood size for $CF_u$ and $CF_i$ to 20.} $sim(G_{u}, G_{v})$ is the cosine similarity between the genre distributions of user $u$ and neighbor $v$. Finally, $|GA_{g, v}|$ indicates how often $v$ has listened to genre $g$ in the past.

\para{Item-based collaborative filtering baseline: \boldmath{$CF_i$}.}
Similar to $CF_u$, $CF_i$ is a collaborative filtering-based approach, but instead of finding similar users for the target user $u$, it aims to find similar items, i.e., music artists $S_{A_u}$, for the artists $A_u$ that $u$ has listened to in the past. Then, it predicts the genres that are assigned to these similar artists as given by:
\begin{equation} \label{eq:cfi}
    \widetilde{G_u^k}~=~\argmax_{g \in G(S_{A_u})}^{k}\left(\sum\limits_{a \in A_{u}}\sum\limits_{s \in S_{a}}{sim(G_{a}, G_{s})}\right)
\end{equation}
where $G(S_{A_u})$ are the genres assigned to the similar artists $S_{A_u}$, $S_{a}$ is the set of similar artists for an artist $a \in A_u$,\footnote{For $A_{u}$, we consider the set of the 20 artists that $u$ has listened to most frequently.} and $sim(G_{a}, G_{s})$ is the cosine similarity between the genre distributions assigned to $a$ and the genres assigned to a similar artist $s \in S_a$.

\para{Popularity-based baseline: \boldmath{$POP_u$}.}
$POP_u$ is a personalized music genre modeling technique, which predicts the $k$ most frequently listened genres in the listening history of user $u$. $POP_u$ is given by the following equation:
\begin{equation} \label{eq:pop}
    \widetilde{G_u^k}~=~\argmax_{g \in G_u}^{k}(|GA_{g, u}|)
\end{equation}
Here, $G_u$ is the set of genres $u$ has listened to in the past and $|GA_{g, u}|$ denotes the number of times $u$ has listened to $g$. Thus, it ranks the genres $u$ has listened to in the past by popularity.

\para{Time-based baseline: \boldmath{$TIME_u$}.}
The time-based baseline $TIME_u$ predicts the $k$ genres that user $u$ has most recently listened to. It is given by:
\begin{equation} \label{eq:time}
    \widetilde{G_u^k}~=~\argmin_{g \in G_u}^{k}(t_{u, g, n})
\end{equation}
where $t_{u, g, n}$ is the time since the last (i.e., the $n$\textsuperscript{th}) LE of $g$ by $u$.

\begin{table*}[t!]
  \setlength{\tabcolsep}{10pt}    
  \centering
    \begin{tabular}{r|r|ccccc|cc}
    \specialrule{.2em}{.1em}{.1em}
User group & Evaluation metric  & $TOP$ & $CF_u$  & $CF_i$  & $POP_u$  & $TIME_u$  & $BLL_u$ & $ACT_{u,a}$ \\\hline 
     \multirow{4}{*}{\centering{LowMS}}                                                                                                                           
    & $F1@5$            & .108 & .311 & .341  & .356    & .368    & .397 & \textbf{.485$^{***}$}   \\
    & $MRR@10$        & .101 & .389  & .425  &    .443        & .445 & .492 & \textbf{.626$^{***}$}    \\
    & $MAP@10$        & .112 & .461  & .505  &    .533        & .550 & .601 & \textbf{.785$^{***}$}    \\
    & $nDCG@10$        &    .180 & .541 & .590   &    .618    & .625 & .679 & \textbf{.824$^{***}$} \\\hline        
                                                                    
    \multirow{4}{*}{\centering{MedMS}}                                                                                                                          
    & $F1@5$            & .196 & .271 & .284    &    .292 & .293  & .338 & \textbf{.502$^{***}$}    \\
    & $MRR@10$        & .146 & .248  & .264     &    .274    & .272  & .320 & \textbf{.511$^{***}$}    \\
    & $MAP@10$        & .187 & .319  & .336     &    .351    & .365  & .419 & \textbf{.705$^{***}$}    \\
    & $nDCG@10$        &    .277 & .419  & .441     &  .460  & .452  & .523 & \textbf{.753$^{***}$}    \\\hline
    
        \multirow{4}{*}{\centering{HighMS}}                                                                                                                          
    & $F1@5$            & .247 & .273 & .266  &    .282    & .228 & .304 & \textbf{.427$^{***}$}   \\
    & $MRR@10$        & .188 & .232  & .229  &    .242        & .201 & .266 & \textbf{.412$^{***}$}    \\
    & $MAP@10$        & .246 & .304  & .298  &    .314        & .267 & .348 & \textbf{.569$^{***}$}   \\
    & $nDCG@10$        &    .354 & .413 & .402  &    .429     & .357 & .462 & \textbf{.642$^{***}$}    \\
    \specialrule{.2em}{.1em}{.1em}                                
    \end{tabular}
    \caption{Genre prediction accuracy results comparing our $BLL_u$ and $ACT_{u,a}$ approaches with a mainstream-based baseline ($TOP$), a user-based collaborative filtering baseline ($CF_u$), an item-based collaborative filtering baseline ($CF_i$), a popularity-based baseline ($POP_u$) and a time-based baseline ($TIME_u$). For all three user groups (i.e., LowMS, MedMS, and HighMS), $ACT_{u,a}$ outperforms all other approaches. According to a t-test with $\alpha$~=~.001, ``$^{***}$'' indicates statistically significant differences between $ACT_{u,a}$ and all other approaches.}    
  \label{tab:results}
\end{table*}

\begin{figure*}[t] 
   \centering
     \captionsetup[subfigure]{justification=centering}
     \subfloat[User group: LowMS]{ 
      \includegraphics[width=0.32\textwidth]{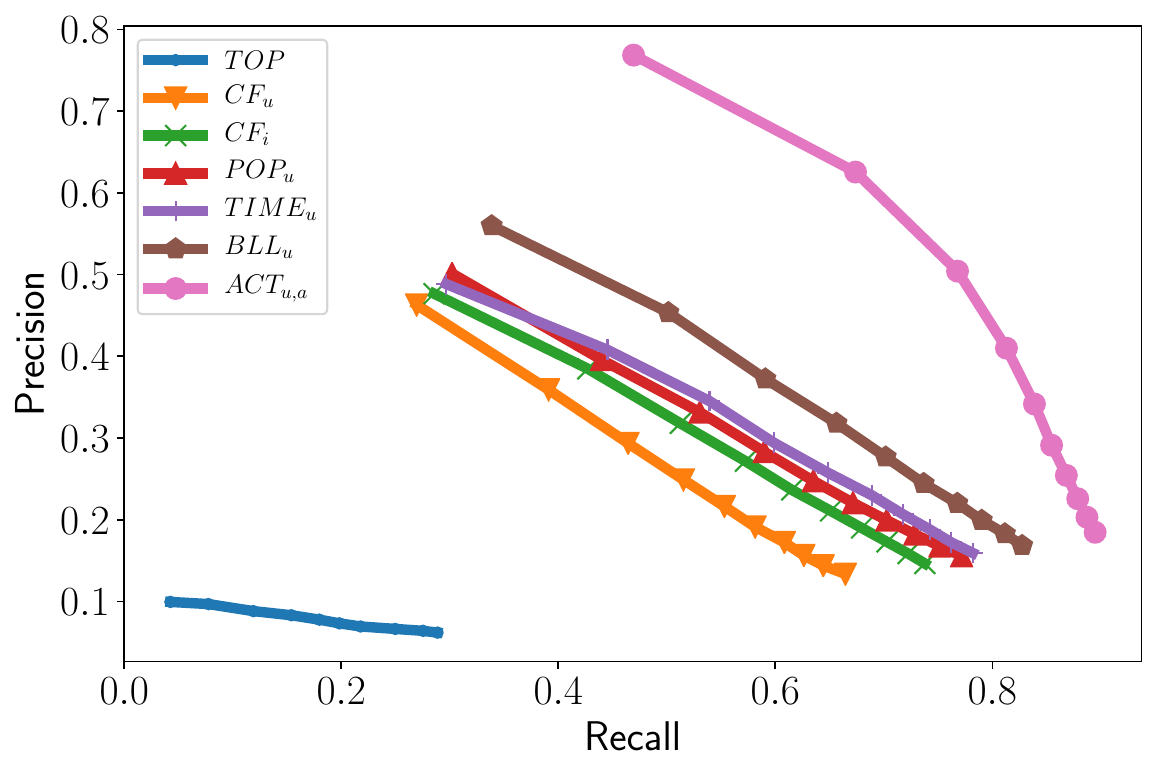}
   }
     \subfloat[User group: MedMS]{ 
      \includegraphics[width=0.32\textwidth]{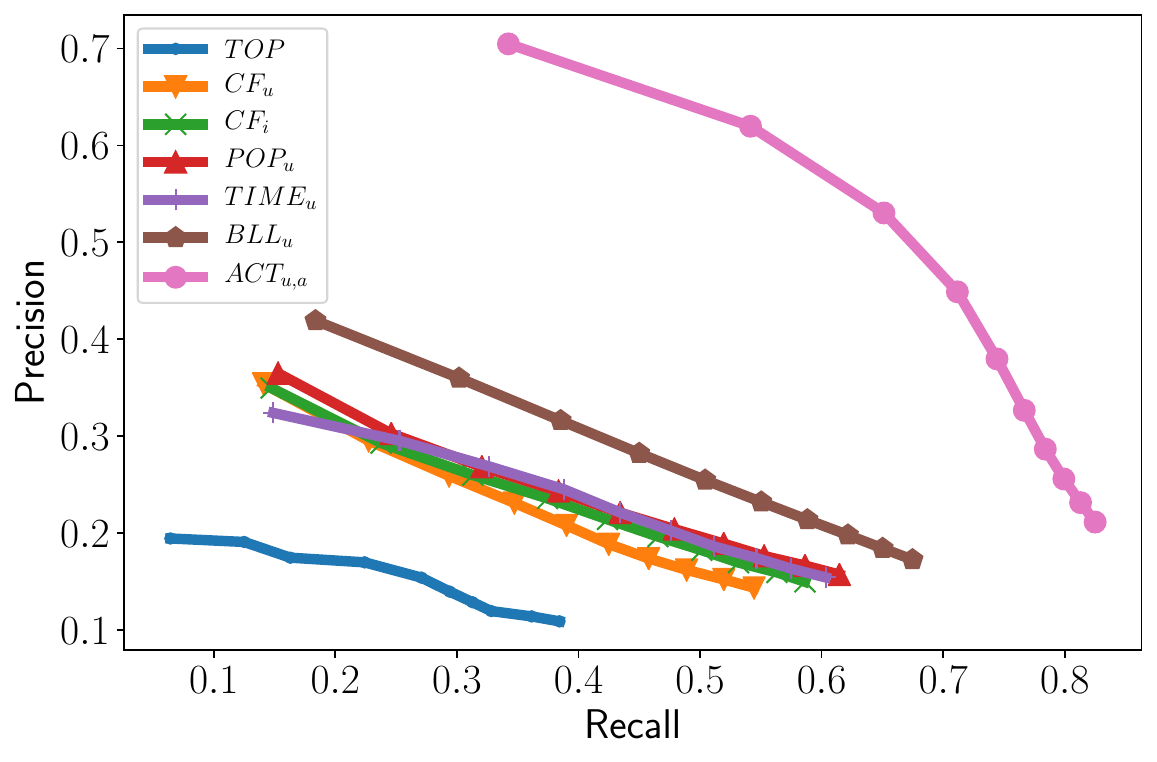} 
   }
     \subfloat[User group: HighMS]{ 
      \includegraphics[width=0.32\textwidth]{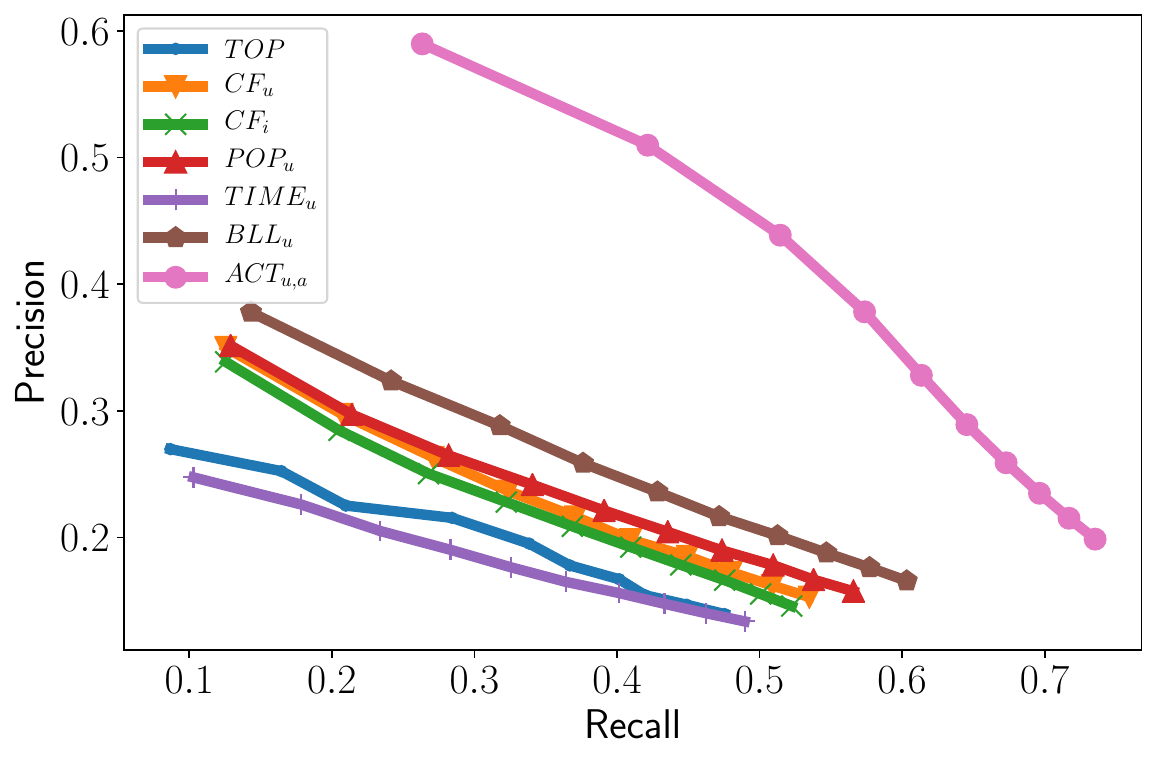} 
   }
   \caption{Recall/precision plots for $k = 1 \ldots 10$ predicted genres of the baselines and our $BLL_u$ and $ACT_{u,a}$ approaches for the three user groups LowMS, MedMS, and HighMS. $ACT_{u,a}$ achieves the best results in all settings.}
     \label{fig:results_plots}
\end{figure*}

\subsection{Evaluation Protocol and Metrics}
We split the datasets into train and test sets~\cite{Cremonesi2008eval}. In doing so, we ensure that our evaluation protocol preserves the temporal order of the LEs, which simulates a real-world scenario in which we predict genres of future LEs based on past ones and not the other way round~\cite{www_hashtag_2017}. This also means that a classic $k$-fold cross-validation evaluation protocol is not useful in our setting.

Specifically, we put the most recent 1\% of the LEs of each user into the test set (i.e., $LE_{test}$) and keep the remaining LEs for the train set (i.e., $LE_{train}$). We do not use a classic 80/20 split as the number of LEs per user is large (i.e., on average, 7,689 LEs per user). Although we only use the most recent  1\% of listening events per user, this process leads to three large test sets with 69,153 listening events for LowMS, 79,007 listening events for MedMS, and 82,510 listening events for HighMS. To finally measure the prediction quality of the approaches, we use the following six well-established performance metrics~\cite{baeza2011modern}:

\para{Recall: $R@k$.}
Recall is calculated as the number of correctly predicted genres divided by the number of relevant genres taken from the LEs in the test set $LE_{test}$. It is a measure for the completeness of the predictions and is formally given by:
  \begin{align}
		R@k = \frac{ 1 }{ |LE_{test}| } \sum\limits_{ u, a \in LE_{test} }{ \frac{|\widetilde{G_{u}^k} \cap G_{u,a}|}{ |G_{u,a}| } }
  \end{align}
where $\widetilde{G_{u}^k}$ denotes the $k$ predicted genres and $G_{u,a}$ the set of relevant genres of an artist $a$ in user $u$'s LEs in the test set.

\para{Precision: $P@k$.}
Precision is calculated as the number of correctly predicted genres divided by the number of predictions $k$ and is a measure for the accuracy of the predictions. It is given by:
  \begin{align} \label{eq:precision}
		P@k = \frac{ 1 }{ |LE_{test}| } \sum\limits_{ u, a \in LE_{test} }{ \frac{|\widetilde{G_{u}^k} \cap G_{u,a}|}{ k } }
  \end{align}
We report recall and precision for $k~=~1 \ldots 10$ predicted genres in form of recall/precision plots.

\para{F1-score: $F1@k$.}
F1-score is the harmonic mean of recall and precision:
  \begin{align}
		F1@k = 2 \cdot \frac{ P@k \cdot R@k }{ P@k + R@k } 
  \end{align}
We report the F1-score for $k$~=~5, where it typically reaches its highest value if 10 genres are predicted.

\para{Mean Reciprocal Rank: MRR@k.}
MRR is the average of reciprocal ranks $r(g)$ of all relevant genres in the list of predicted genres:
	\begin{align}
		MRR@k = \frac{ 1 }{ |LE_{test}| } \sum\limits_{ u, a \in LE_{test} }{ \frac{ 1 }{ |G_{u,a}| } \sum\limits_{ g \in G_{u,a} }{ \frac{ 1 }{ r(g) } } }
	\end{align}
This means that a high MRR is achieved if relevant genres occur at the beginning of the predicted genre list.

\para{Mean Average Precision: MAP@k.}
MAP is an extension of the precision metric by also taking the ranking of the correctly predicted genres into account and is given by:
	\begin{align}
		MAP@k = \frac{ 1 }{ |LE_{test}| } \sum\limits_{ u, a \in LE_{test} }{ \frac{ 1 }{ |G_{u,a}| } \sum\limits_{ i = 1 }\limits^{ k }{ Rel_i \cdot P@i } }
	\end{align}
Here, $Rel_i$ is 1 if the predicted genre at position $i$ is among the relevant genres (0 otherwise) and $P@i$ is the precision calculated at position $i$ according to Equation~\ref{eq:precision}.

\para{Normalized Discounted Cumulative Gain: nDCG@k.}
nDCG is another ranking-dependent metric. It is based on the Discounted Cumulative Gain ($DCG@k$) measure~\cite{jarvelin2008discounted}, which is defined as:
   \begin{align}
      DCG@k = \sum\limits_{i = 1}\limits^{k} \left(\frac{2 ^ {Rel_i} - 1}{log_{2}(1 + i)}\right)
   \end{align}
where $Rel_i$ is 1 if the genre predicted for the $i^{th}$ item is relevant (0 otherwise). $nDCG@k$ is given as $DCG@k$ divided by $iDCG@k$, which is the highest possible DCG value that can be achieved if all relevant genres are predicted in the correct order:
  \begin{align}
      nDCG@k = \frac{ 1 }{ |LE_{test}| } \sum\limits_{ u, a \in LE_{test} }{ \left(\frac{DCG@k}{iDCG@k}\right) }
\end{align}   
We report MRR, MAP, and nDCG for $k~=~10$ predicted music genres, where these metrics reach their highest values.

\subsection{Results and Discussion}
In this section, we present and discuss our evaluation results. The accuracy results according to $F1@5$, $MRR@10$, $MAP@10$, and $nDCG@10$ are shown in Table~\ref{tab:results} for the five baseline approaches as well as the proposed $BLL_u$ and $ACT_{u,a}$ algorithms. Furthermore, we provide recall/precision plots for $k = 1 \ldots 10$ predicted genres.

\para{Accuracy of baseline approaches.}
When analyzing the performance of the baseline approaches $TOP$, $CF_u$, $CF_i$, $POP_u$, and $TIME_u$, we see a clear difference between the non personalized and the personalized algorithms. While the non personalized $TOP$ approach, which predicts the top-$k$ genres of the mainstream, provides better accuracy results in the HighMS setting than in the LowMS setting, the personalized $CF_u$, $CF_i$, $POP_u$, and $TIME_u$ algorithms provide better results in the LowMS setting than in the HighMS setting. Hence, personalized genre modeling approaches provide better results, the lower the mainstreaminess of the users. Non-personalized genre modeling approaches, however, have higher performance, the higher the mainstreaminess of the users.

\begin{figure}[t]
   \centering 
   \includegraphics[width=0.48\textwidth]{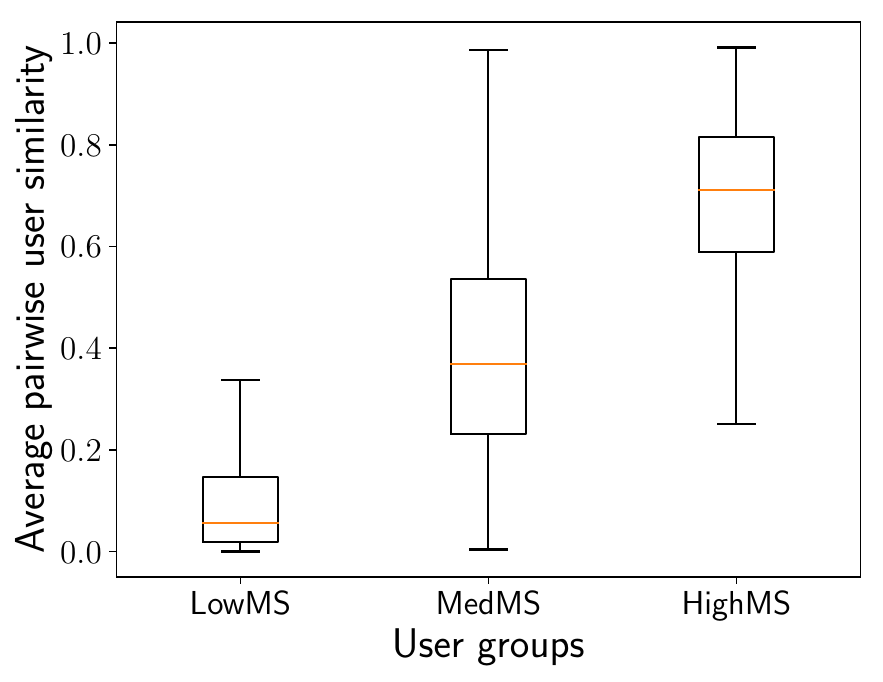} 
     \caption{Average pairwise user similarity for LowMS, MedMS, and HighMS. We calculate the user similarity using the cosine similarity metric based on the users' genre distributions. While users in the LowMS group show a very individual listening behavior, users in the HighMS group tend to listen to similar music genres.}
     \label{fig:user_sim}
\end{figure}

Next, we compare the accuracy of the two collaborative filtering-based methods, $CF_u$, and $CF_i$. Here, the item-based CF variant $CF_i$ reaches higher accuracy estimates in the LowMS and MedMS settings, while the user-based CF variant $CF_u$ provides better performance in the HighMS setting. To better understand this pattern of results, we provide the average pairwise user similarity in the form of boxplots in Figure~\ref{fig:user_sim}. Here, for all three user groups, we calculate the pairwise similarity between the users via the cosine similarity metric based on the users' genre distribution vectors. We see that users in the HighMS setting are very similar to each other, which explains the good performance of an algorithm that is based on user similarities, such as $CF_u$.

$ POP_u$ and $TIME_u$ reach the highest accuracy estimates among the five baseline approaches. Interestingly, the popularity-based $POP_u$ algorithm provides the best results for the HighMS user group, while the time-based $TIME_u$ algorithm provides the best results in the LowMS user group. For the MedMS user group, however, both algorithms reach a comparable accuracy performance, which shows the importance of both factors, frequency (i.e., popularity) and recency (i.e., time).

\para{Accuracy of $BLL_u$ and $ACT_{u,a}$.}
We discuss the results of the $BLL_u$ and $ACT_{u,a}$ approaches, which utilize human memory processes as defined by the cognitive architecture ACT-R in order to model and predict music genre preferences. Specifically, $BLL_u$ combines the factors of past usage frequency and recency via the BLL equation (see Equation~\ref{eq:bllu}) and $ACT_{u,a}$ extends $BLL_u$ by also considering the current context via the activation equation (see Equation~\ref{eq:activation}). In this work, we define the current context by the genres assigned to the artist that the target user $u$ has listened to most recently.

\begin{figure}[t]
   \centering 
   \includegraphics[width=0.48\textwidth]{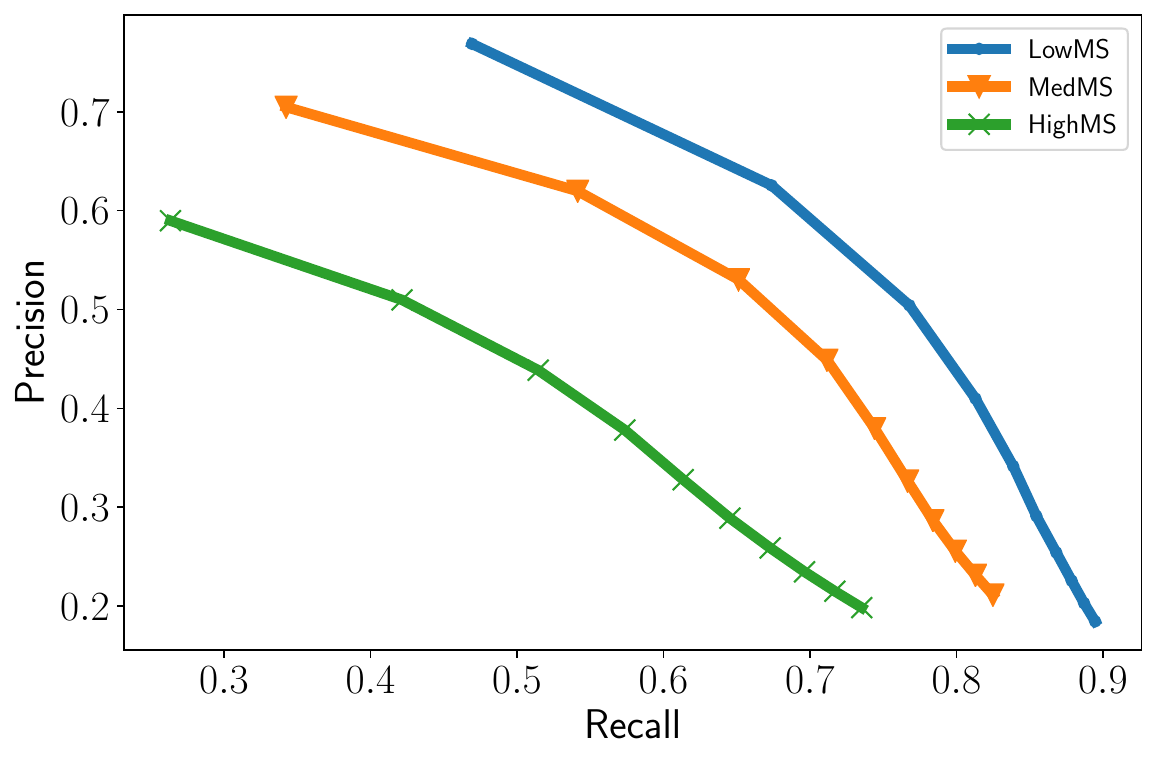} 
     \caption{Recall/precision plot of our $ACT_{u,a}$ approach for $k~=~1 \ldots 10$ predicted genres for the three user groups LowMS, MedMS, and HighMS. We observe good prediction accuracy results for $ACT_{u,a}$ in all settings but especially for LowMS. This shows that our approach based on human memory processes is especially useful for predicting the music genre preferences of LowMS users.\vspace{-3mm}}
     \label{fig:bll_plot}
\end{figure}

As expected, when combining the factors of past usage frequency and recency in the form of $BLL_u$, we can outperform the best performing baseline approaches $POP_u$ and $TIME_u$ in all three settings (i.e., LowMS, MedMS, and HighMS). We can further improve the accuracy performance when we additionally consider the current context in the form of $ACT_{u,a}$. Here, we reach a statistically significant improvement\footnote{According to a t-test with $\alpha$ = .001.} over all other approaches across all evaluation metrics and user groups. Furthermore, in Figure~\ref{fig:bll_plot}, we present a recall/precision plot showing the accuracy of $ACT_{u,a}$ for $k~=~1 \ldots 10$ predicted genres for LowMS, MedMS, and HighMS. We observe good results for all three user groups but especially in the LowMS setting, in which we are faced with users with a low interest in mainstream music.

This shows that the proposed $ACT_{u,a}$ algorithm can provide accurate predictions of music genres listened to in the future for all user groups and, thus, treats all users in our experiment in a fair manner.
Moreover, since our approach utilizes human memory processes, it is based on psychological principles of human intelligence rather than artificial intelligence. We believe that this theoretical underpinning contributes to the explanation effectiveness of our approach as we can fully understand why a specific genre was predicted for a target user in a given context. To further illustrate this with an example, we would like to refer back to Figure~\ref{fig:example}. In this figure, we have shown the differences between $BLL_u$ and $ACT_{u,a}$ for two predicted genres $g_1$ and $g_2$. Let us assume that these are the top-$2$ predicted genres for a target user $u$. According to $BLL_u$, we know that these genres got the highest activation levels because $u$ has listened to them very frequently and recently. When looking at the activation levels calculated by $ACT_{u,a}$, we also take the current context into account and, thus, get an indication for the similarity of $g_1$ and $g_2$ to the genres assigned to the most recently listened artist $a$ of user $u$. In our example, genre $g_2$ is strongly related to the current context, while genre $g_1$ only has a weak relation to it. Taken together, with our $ACT_{u,a}$ approach, we can easily explain genre prediction results according to three simple factors that are relevant for human memory processes according to the cognitive architecture ACT-R: (i) past usage frequency, (ii) past usage recency, and (iii) similarity to current context.

\section{Conclusion and Future Work}
\label{s:conc}
In this paper, we presented $BLL_u$ and $ACT_{u,a}$, two music genre preference modeling, and prediction approaches based on the human memory module of the cognitive architecture ACT-R. While $BLL_u$ utilizes the BLL equation of ACT-R in order to model the factors of past usage frequency (i.e., popularity) and recency (i.e., time), $ACT_{u,a}$ integrates the activation equation of ACT-R to also incorporate the current context. We defined this context as the genres assigned to the most recently listened artist of the target user. 
Using a dataset gathered from the music platform Last.fm, we evaluated $BLL_u$ and $ACT_{u,a}$ against a mainstream-based approach $TOP$, a user-based CF approach $CF_u$, an item-based CF approach $CF_i$, a popularity-based approach $POP_u$ as well as a time-based approach $TIME_u$. We used six evaluation metrics (i.e., recall, precision, F1-score, MRR, MAP, and nDCG) in three evaluation settings in which the evaluated users differed in terms of their inclination to mainstream music (i.e., LowMS, MedMS, and HighMS user groups). Our evaluation results show that both $BLL_u$ and $ACT_{u,a}$ outperform the five baseline methods in all three settings; $ACT_{u,a}$ even does so in a statistically significant manner. Furthermore, we find that especially the current context is of high importance when aiming for accurate genre predictions.

Summed up, in this work, we have shown that human memory processes in the form of ACT-R's activation equation can be effectively utilized for modeling and predicting music genres. By following such a psychology-inspired approach, we also believe that we can model a user's preferences transparently, in contrast to, e.g., deep learning-based approaches based on latent user representations. Therefore, our approach could be useful to realize more transparent and explainable music recommender systems.

In the present work, we only considered the genres assigned to the most recently listened artist of the target user as contextual information. However, related work on music preference modeling has shown that music listening habits depend on the time of the day, the current activity of a user or the mood a user is currently experiencing (see, e.g.,~\cite{knees2019awareness}). 
For future work, we also plan to utilize the procedural memory processes of ACT-R in addition to the activation equation. As, for instance, done in the SNIF-ACT model~\cite{pirolli2003snif,fu2007snif}, we could define so-called production rules in order to transfer the user's preferences into actual music recommendation strategies. By making these rules transparent to the user, we aim to contribute to research on transparent recommender systems that create explainable recommendations.

To foster the reproducibility of our research, we use the publicly available LFM-1b dataset (see Section~\ref{s:method}). Furthermore, we provide the source code of our approach as part of our TagRec framework~\cite{kowald2017tagrec}.

\balance{}
\bibliographystyle{ACM-Reference-Format}
\bibliography{document}


\begin{thebibliography}{35}


\ifx \showCODEN    \undefined \def \showCODEN     #1{\unskip}     \fi
\ifx \showDOI      \undefined \def \showDOI       #1{#1}\fi
\ifx \showISBNx    \undefined \def \showISBNx     #1{\unskip}     \fi
\ifx \showISBNxiii \undefined \def \showISBNxiii  #1{\unskip}     \fi
\ifx \showISSN     \undefined \def \showISSN      #1{\unskip}     \fi
\ifx \showLCCN     \undefined \def \showLCCN      #1{\unskip}     \fi
\ifx \shownote     \undefined \def \shownote      #1{#1}          \fi
\ifx \showarticletitle \undefined \def \showarticletitle #1{#1}   \fi
\ifx \showURL      \undefined \def \showURL       {\relax}        \fi
\providecommand\bibfield[2]{#2}
\providecommand\bibinfo[2]{#2}
\providecommand\natexlab[1]{#1}
\providecommand\showeprint[2][]{arXiv:#2}

\bibitem[\protect\citeauthoryear{Anderson, Bothell, Byrne, Douglass, Lebiere,
  and Qin}{Anderson et~al\mbox{.}}{2004}]%
        {anderson2004integrated}
\bibfield{author}{\bibinfo{person}{John~R Anderson}, \bibinfo{person}{Daniel
  Bothell}, \bibinfo{person}{Michael~D Byrne}, \bibinfo{person}{Scott
  Douglass}, \bibinfo{person}{Christian Lebiere}, {and} \bibinfo{person}{Yulin
  Qin}.} \bibinfo{year}{2004}\natexlab{}.
\newblock \showarticletitle{An integrated theory of the mind}.
\newblock \bibinfo{journal}{\emph{Psychological review}} \bibinfo{volume}{111},
  \bibinfo{number}{4} (\bibinfo{year}{2004}), \bibinfo{pages}{25 pages}.
\newblock


\bibitem[\protect\citeauthoryear{Anderson and Schooler}{Anderson and
  Schooler}{1991}]%
        {anderson1991reflections}
\bibfield{author}{\bibinfo{person}{John~R Anderson} {and}
  \bibinfo{person}{Lael~J Schooler}.} \bibinfo{year}{1991}\natexlab{}.
\newblock \showarticletitle{Reflections of the environment in memory}.
\newblock \bibinfo{journal}{\emph{Psychological science}} \bibinfo{volume}{2},
  \bibinfo{number}{6} (\bibinfo{year}{1991}), \bibinfo{pages}{396--408}.
\newblock


\bibitem[\protect\citeauthoryear{Baeza-Yates, Ribeiro,
  et~al\mbox{.}}{Baeza-Yates et~al\mbox{.}}{2011}]%
        {baeza2011modern}
\bibfield{author}{\bibinfo{person}{Ricardo Baeza-Yates},
  \bibinfo{person}{Berthier de Ara{\'u}jo~Neto Ribeiro}, {et~al\mbox{.}}}
  \bibinfo{year}{2011}\natexlab{}.
\newblock \bibinfo{booktitle}{\emph{Modern Information Retrieval}}.
\newblock \bibinfo{publisher}{New York: ACM Press; Harlow, England:
  Addison-Wesley}.
\newblock


\bibitem[\protect\citeauthoryear{Bauer and Schedl}{Bauer and Schedl}{2019}]%
        {bauer2019global}
\bibfield{author}{\bibinfo{person}{Christine Bauer} {and}
  \bibinfo{person}{Markus Schedl}.} \bibinfo{year}{2019}\natexlab{}.
\newblock \showarticletitle{Global and country-specific mainstreaminess
  measures: Definitions, analysis, and usage for improving personalized music
  recommendation systems}.
\newblock \bibinfo{journal}{\emph{PloS one}} \bibinfo{volume}{14},
  \bibinfo{number}{6} (\bibinfo{year}{2019}), \bibinfo{pages}{e0217389}.
\newblock


\bibitem[\protect\citeauthoryear{Cantor and Zillmann}{Cantor and
  Zillmann}{1973}]%
        {cantor1973effect}
\bibfield{author}{\bibinfo{person}{Joanne~R Cantor} {and} \bibinfo{person}{Dolf
  Zillmann}.} \bibinfo{year}{1973}\natexlab{}.
\newblock \showarticletitle{The effect of affective state and emotional arousal
  on music appreciation}.
\newblock \bibinfo{journal}{\emph{The Journal of General Psychology}}
  \bibinfo{volume}{89}, \bibinfo{number}{1} (\bibinfo{year}{1973}),
  \bibinfo{pages}{97--108}.
\newblock


\bibitem[\protect\citeauthoryear{Cremonesi, Turrin, Lentini, and
  Matteucci}{Cremonesi et~al\mbox{.}}{2008}]%
        {Cremonesi2008eval}
\bibfield{author}{\bibinfo{person}{Paolo Cremonesi}, \bibinfo{person}{Roberto
  Turrin}, \bibinfo{person}{Eugenio Lentini}, {and} \bibinfo{person}{Matteo
  Matteucci}.} \bibinfo{year}{2008}\natexlab{}.
\newblock \showarticletitle{An Evaluation Methodology for Collaborative
  Recommender Systems}. In \bibinfo{booktitle}{\emph{Proceedings of
  AXMEDIS'2008}}. \bibinfo{publisher}{IEEE Computer Society},
  \bibinfo{address}{Washington, DC, USA}, \bibinfo{pages}{224--231}.
\newblock
\showISBNx{978-0-7695-3406-0}
\urldef\tempurl%
\url{https://doi.org/10.1109/AXMEDIS.2008.13}
\showDOI{\tempurl}


\bibitem[\protect\citeauthoryear{Fu and Pirolli}{Fu and Pirolli}{2007}]%
        {fu2007snif}
\bibfield{author}{\bibinfo{person}{Wai-Tat Fu} {and} \bibinfo{person}{Peter
  Pirolli}.} \bibinfo{year}{2007}\natexlab{}.
\newblock \showarticletitle{SNIF-ACT: A cognitive model of user navigation on
  the World Wide Web}.
\newblock \bibinfo{journal}{\emph{Human--Computer Interaction}}
  \bibinfo{volume}{22}, \bibinfo{number}{4} (\bibinfo{year}{2007}),
  \bibinfo{pages}{355--412}.
\newblock


\bibitem[\protect\citeauthoryear{J{\"a}rvelin, Price, Delcambre, and
  Nielsen}{J{\"a}rvelin et~al\mbox{.}}{2008}]%
        {jarvelin2008discounted}
\bibfield{author}{\bibinfo{person}{Kalervo J{\"a}rvelin},
  \bibinfo{person}{Susan~L Price}, \bibinfo{person}{Lois~ML Delcambre}, {and}
  \bibinfo{person}{Marianne~Lykke Nielsen}.} \bibinfo{year}{2008}\natexlab{}.
\newblock \showarticletitle{Discounted cumulated gain based evaluation of
  multiple-query IR sessions}. In \bibinfo{booktitle}{\emph{Proceedings of
  ECIR'2008}}. \bibinfo{publisher}{Springer}, \bibinfo{pages}{4--15}.
\newblock


\bibitem[\protect\citeauthoryear{Juslin and Sloboda}{Juslin and
  Sloboda}{2001}]%
        {juslin2001music}
\bibfield{author}{\bibinfo{person}{Patrik~N Juslin} {and}
  \bibinfo{person}{John~A Sloboda}.} \bibinfo{year}{2001}\natexlab{}.
\newblock \bibinfo{booktitle}{\emph{Music and emotion: Theory and research.}}
\newblock \bibinfo{publisher}{Oxford University Press}.
\newblock


\bibitem[\protect\citeauthoryear{Knees, Schedl, Ferwerda, and Laplante}{Knees
  et~al\mbox{.}}{2019}]%
        {knees2019awareness}
\bibfield{author}{\bibinfo{person}{P. Knees}, \bibinfo{person}{M. Schedl},
  \bibinfo{person}{B. Ferwerda}, {and} \bibinfo{person}{A. Laplante}.}
  \bibinfo{year}{2019}\natexlab{}.
\newblock \showarticletitle{User Awareness in Music Recommender Systems}. In
  \bibinfo{booktitle}{\emph{Mirjam Augstein, Eelco Herder, Wolfgang Wörndl
  (eds.), Personalized Human-Computer Interaction}}. DeGruyter.
\newblock


\bibitem[\protect\citeauthoryear{Kowald, Kopeinik, and Lex}{Kowald
  et~al\mbox{.}}{2017a}]%
        {kowald2017tagrec}
\bibfield{author}{\bibinfo{person}{Dominik Kowald}, \bibinfo{person}{Simone
  Kopeinik}, {and} \bibinfo{person}{Elisabeth Lex}.}
  \bibinfo{year}{2017}\natexlab{a}.
\newblock \showarticletitle{The tagrec framework as a toolkit for the
  development of tag-based recommender systems}. In
  \bibinfo{booktitle}{\emph{Adjunct Publication of UMAP'2017}}.
  \bibinfo{publisher}{ACM}, \bibinfo{pages}{23--28}.
\newblock


\bibitem[\protect\citeauthoryear{Kowald and Lex}{Kowald and Lex}{2016}]%
        {Kowald2016bllhypertext}
\bibfield{author}{\bibinfo{person}{Dominik Kowald} {and}
  \bibinfo{person}{Elisabeth Lex}.} \bibinfo{year}{2016}\natexlab{}.
\newblock \showarticletitle{The Influence of Frequency, Recency and Semantic
  Context on the Reuse of Tags in Social Tagging Systems}. In
  \bibinfo{booktitle}{\emph{Proceedings of Hypertext'2016}}.
  \bibinfo{publisher}{ACM}, \bibinfo{address}{New York, NY, USA},
  \bibinfo{pages}{237--242}.
\newblock
\showISBNx{978-1-4503-4247-6}


\bibitem[\protect\citeauthoryear{Kowald, Lex, and Schedl}{Kowald
  et~al\mbox{.}}{2019}]%
        {ismir_lfm_2019}
\bibfield{author}{\bibinfo{person}{Dominik Kowald}, \bibinfo{person}{Elisabeth
  Lex}, {and} \bibinfo{person}{Markus Schedl}.}
  \bibinfo{year}{2019}\natexlab{}.
\newblock \showarticletitle{Modeling Artist Preferences for Personalized Music
  Recommendations}. In \bibinfo{booktitle}{\emph{Proc. of ISMIR '19}}.
\newblock


\bibitem[\protect\citeauthoryear{Kowald, Pujari, and Lex}{Kowald
  et~al\mbox{.}}{2017b}]%
        {www_hashtag_2017}
\bibfield{author}{\bibinfo{person}{Dominik Kowald},
  \bibinfo{person}{Subhash~Chandra Pujari}, {and} \bibinfo{person}{Elisabeth
  Lex}.} \bibinfo{year}{2017}\natexlab{b}.
\newblock \showarticletitle{Temporal Effects on Hashtag Reuse in Twitter: A
  Cognitive-Inspired Hashtag Recommendation Approach}. In
  \bibinfo{booktitle}{\emph{Proceedings of WWW'2017}}.
  \bibinfo{publisher}{ACM}, \bibinfo{pages}{10 pages}.
\newblock


\bibitem[\protect\citeauthoryear{Kowald, Schedl, and Lex}{Kowald
  et~al\mbox{.}}{2020}]%
        {kowald2020unfairness}
\bibfield{author}{\bibinfo{person}{Dominik Kowald}, \bibinfo{person}{Markus
  Schedl}, {and} \bibinfo{person}{Elisabeth Lex}.}
  \bibinfo{year}{2020}\natexlab{}.
\newblock \showarticletitle{The unfairness of popularity bias in music
  recommendation: A reproducibility study}. In \bibinfo{booktitle}{\emph{Proc.
  of ECIR'20}}. Springer, \bibinfo{pages}{35--42}.
\newblock


\bibitem[\protect\citeauthoryear{North and Hargreaves}{North and
  Hargreaves}{2008}]%
        {north2008social}
\bibfield{author}{\bibinfo{person}{Adrian North} {and} \bibinfo{person}{David
  Hargreaves}.} \bibinfo{year}{2008}\natexlab{}.
\newblock \bibinfo{booktitle}{\emph{The social and applied psychology of
  music}}.
\newblock \bibinfo{publisher}{OUP Oxford}.
\newblock


\bibitem[\protect\citeauthoryear{Oord, Dieleman, and Schrauwen}{Oord
  et~al\mbox{.}}{2013}]%
        {Oord2013:DCM}
\bibfield{author}{\bibinfo{person}{A\"{a}ron van~den Oord},
  \bibinfo{person}{Sander Dieleman}, {and} \bibinfo{person}{Benjamin
  Schrauwen}.} \bibinfo{year}{2013}\natexlab{}.
\newblock \showarticletitle{Deep Content-based Music Recommendation}. In
  \bibinfo{booktitle}{\emph{Proceedings of NIPS'2013}}.
  \bibinfo{publisher}{Curran Associates Inc.}, \bibinfo{address}{USA},
  \bibinfo{pages}{2643--2651}.
\newblock


\bibitem[\protect\citeauthoryear{Pereira, Teixeira, Figueiredo, Xavier, Castro,
  and Brattico}{Pereira et~al\mbox{.}}{2011}]%
        {pereira2011music}
\bibfield{author}{\bibinfo{person}{Carlos~Silva Pereira},
  \bibinfo{person}{Jo{\~a}o Teixeira}, \bibinfo{person}{Patr{\'\i}cia
  Figueiredo}, \bibinfo{person}{Jo{\~a}o Xavier},
  \bibinfo{person}{S{\~a}o~Lu{\'\i}s Castro}, {and} \bibinfo{person}{Elvira
  Brattico}.} \bibinfo{year}{2011}\natexlab{}.
\newblock \showarticletitle{Music and emotions in the brain: familiarity
  matters}.
\newblock \bibinfo{journal}{\emph{PloS one}} \bibinfo{volume}{6},
  \bibinfo{number}{11} (\bibinfo{year}{2011}), \bibinfo{pages}{e27241}.
\newblock


\bibitem[\protect\citeauthoryear{Pirolli and Fu}{Pirolli and Fu}{2003}]%
        {pirolli2003snif}
\bibfield{author}{\bibinfo{person}{Peter Pirolli} {and}
  \bibinfo{person}{Wai-Tat Fu}.} \bibinfo{year}{2003}\natexlab{}.
\newblock \showarticletitle{SNIF-ACT: A model of information foraging on the
  World Wide Web}. In \bibinfo{booktitle}{\emph{International Conference on
  User Modeling}}. \bibinfo{publisher}{Springer}, \bibinfo{pages}{45--54}.
\newblock


\bibitem[\protect\citeauthoryear{Rentfrow and Gosling}{Rentfrow and
  Gosling}{2003}]%
        {rentfrow2003re}
\bibfield{author}{\bibinfo{person}{Peter~J Rentfrow} {and}
  \bibinfo{person}{Samuel~D Gosling}.} \bibinfo{year}{2003}\natexlab{}.
\newblock \showarticletitle{The do re mi's of everyday life: the structure and
  personality correlates of music preferences}.
\newblock \bibinfo{journal}{\emph{Journal of personality and social
  psychology}} \bibinfo{volume}{84}, \bibinfo{number}{6}
  (\bibinfo{year}{2003}), \bibinfo{pages}{21 pages}.
\newblock


\bibitem[\protect\citeauthoryear{Schedl}{Schedl}{2016}]%
        {schedl2016lfm}
\bibfield{author}{\bibinfo{person}{Markus Schedl}.}
  \bibinfo{year}{2016}\natexlab{}.
\newblock \showarticletitle{{The LFM-1b Dataset for Music Retrieval and
  Recommendation}}. In \bibinfo{booktitle}{\emph{Proceedings of the 2016
  Conference on Multimedia Retrieval}}. \bibinfo{publisher}{ACM},
  \bibinfo{pages}{103--110}.
\newblock


\bibitem[\protect\citeauthoryear{Schedl and Bauer}{Schedl and Bauer}{2017}]%
        {schedl2017distance}
\bibfield{author}{\bibinfo{person}{Markus Schedl} {and}
  \bibinfo{person}{Christine Bauer}.} \bibinfo{year}{2017}\natexlab{}.
\newblock \showarticletitle{Distance-and Rank-based Music Mainstreaminess
  Measurement}. In \bibinfo{booktitle}{\emph{Adjunct Publication of the 25th
  Conference on User Modeling, Adaptation and Personalization}}. ACM,
  \bibinfo{pages}{364--367}.
\newblock


\bibitem[\protect\citeauthoryear{Schedl and Bauer}{Schedl and Bauer}{2018}]%
        {schedl_bauer:jmm:2018}
\bibfield{author}{\bibinfo{person}{Markus Schedl} {and}
  \bibinfo{person}{Christine Bauer}.} \bibinfo{year}{2018}\natexlab{}.
\newblock \showarticletitle{{An Analysis of Global and Regional Mainstreaminess
  for Personalized Music Recommender Systems}}.
\newblock \bibinfo{journal}{\emph{Journal of Mobile Multimedia}}
  \bibinfo{volume}{14} (\bibinfo{year}{2018}), \bibinfo{pages}{95--112}.
\newblock


\bibitem[\protect\citeauthoryear{Schedl and Ferwerda}{Schedl and
  Ferwerda}{2017}]%
        {schedl2017large}
\bibfield{author}{\bibinfo{person}{Markus Schedl} {and} \bibinfo{person}{Bruce
  Ferwerda}.} \bibinfo{year}{2017}\natexlab{}.
\newblock \showarticletitle{{Large-scale Analysis of Group-specific Music Genre
  Taste From Collaborative Tags}}. In \bibinfo{booktitle}{\emph{Proceedings of
  ISM'2017}}. \bibinfo{publisher}{IEEE}, \bibinfo{pages}{479--482}.
\newblock


\bibitem[\protect\citeauthoryear{Schedl, G{\'o}mez, Trent, Tkal\v{c}i\v{c},
  Eghbal-Zadeh, and Martorell}{Schedl et~al\mbox{.}}{2018a}]%
        {schedl_etal:taffc:2017}
\bibfield{author}{\bibinfo{person}{Markus Schedl}, \bibinfo{person}{Emilia
  G{\'o}mez}, \bibinfo{person}{Erika Trent}, \bibinfo{person}{Marko
  Tkal\v{c}i\v{c}}, \bibinfo{person}{Hamid Eghbal-Zadeh}, {and}
  \bibinfo{person}{Agust\'in Martorell}.} \bibinfo{year}{2018}\natexlab{a}.
\newblock \showarticletitle{{On the Interrelation between Listener
  Characteristics and the Perception of Emotions in Classical Orchestra
  Music}}.
\newblock \bibinfo{journal}{\emph{IEEE Transactions on Affective Computing}}
  \bibinfo{volume}{9} (\bibinfo{year}{2018}), \bibinfo{pages}{507--525}.
\newblock
Issue 4.


\bibitem[\protect\citeauthoryear{Schedl and Hauger}{Schedl and Hauger}{2015}]%
        {schedl2015tailoring}
\bibfield{author}{\bibinfo{person}{Markus Schedl} {and} \bibinfo{person}{David
  Hauger}.} \bibinfo{year}{2015}\natexlab{}.
\newblock \showarticletitle{Tailoring music recommendations to users by
  considering diversity, mainstreaminess, and novelty}. In
  \bibinfo{booktitle}{\emph{Proceedings of SIGIR'2015}}.
  \bibinfo{publisher}{ACM}, \bibinfo{pages}{947--950}.
\newblock


\bibitem[\protect\citeauthoryear{Schedl, Knees, McFee, Bogdanov, and
  Kaminskas}{Schedl et~al\mbox{.}}{2015}]%
        {schedl2015music}
\bibfield{author}{\bibinfo{person}{Markus Schedl}, \bibinfo{person}{Peter
  Knees}, \bibinfo{person}{Brian McFee}, \bibinfo{person}{Dmitry Bogdanov},
  {and} \bibinfo{person}{Marius Kaminskas}.} \bibinfo{year}{2015}\natexlab{}.
\newblock \showarticletitle{Music recommender systems}.
\newblock In \bibinfo{booktitle}{\emph{Recommender systems handbook}}.
  \bibinfo{publisher}{Springer}, \bibinfo{pages}{453--492}.
\newblock


\bibitem[\protect\citeauthoryear{Schedl, Zamani, Chen, Deldjoo, and
  Elahi}{Schedl et~al\mbox{.}}{2018b}]%
        {Schedl2018challengesmrs}
\bibfield{author}{\bibinfo{person}{Markus Schedl}, \bibinfo{person}{Hamed
  Zamani}, \bibinfo{person}{Ching-Wei Chen}, \bibinfo{person}{Yashar Deldjoo},
  {and} \bibinfo{person}{Mehdi Elahi}.} \bibinfo{year}{2018}\natexlab{b}.
\newblock \showarticletitle{Current challenges and visions in music recommender
  systems research}.
\newblock \bibinfo{journal}{\emph{International Journal of Multimedia
  Information Retrieval}} \bibinfo{volume}{7}, \bibinfo{number}{2}
  (\bibinfo{date}{01 Jun} \bibinfo{year}{2018}), \bibinfo{pages}{95--116}.
\newblock


\bibitem[\protect\citeauthoryear{Schubert}{Schubert}{2007}]%
        {schubert2007influence}
\bibfield{author}{\bibinfo{person}{Emery Schubert}.}
  \bibinfo{year}{2007}\natexlab{}.
\newblock \showarticletitle{The influence of emotion, locus of emotion and
  familiarity upon preference in music}.
\newblock \bibinfo{journal}{\emph{Psychology of Music}} \bibinfo{volume}{35},
  \bibinfo{number}{3} (\bibinfo{year}{2007}), \bibinfo{pages}{499--515}.
\newblock


\bibitem[\protect\citeauthoryear{Shi, Larson, and Hanjalic}{Shi
  et~al\mbox{.}}{2014}]%
        {shi_etal:compsurv:2014}
\bibfield{author}{\bibinfo{person}{Yue Shi}, \bibinfo{person}{Martha Larson},
  {and} \bibinfo{person}{Alan Hanjalic}.} \bibinfo{year}{2014}\natexlab{}.
\newblock \showarticletitle{{Collaborative Filtering Beyond the User-Item
  Matrix: A Survey of the State of the Art and Future Challenges}}.
\newblock \bibinfo{journal}{\emph{Comput. Surveys}} \bibinfo{volume}{47},
  \bibinfo{number}{1}, Article \bibinfo{articleno}{3} (\bibinfo{date}{May}
  \bibinfo{year}{2014}), \bibinfo{numpages}{45}~pages.
\newblock
\showISSN{0360-0300}


\bibitem[\protect\citeauthoryear{Trattner, Kowald, Seitlinger, Kopeinik, and
  Ley}{Trattner et~al\mbox{.}}{2016}]%
        {trattner2016modeling}
\bibfield{author}{\bibinfo{person}{Christoph Trattner},
  \bibinfo{person}{Dominik Kowald}, \bibinfo{person}{Paul Seitlinger},
  \bibinfo{person}{Simone Kopeinik}, {and} \bibinfo{person}{Tobias Ley}.}
  \bibinfo{year}{2016}\natexlab{}.
\newblock \showarticletitle{Modeling Activation Processes in Human Memory to
  Predict the Reuse of Tags}.
\newblock \bibinfo{journal}{\emph{The Journal of Web Science}}
  \bibinfo{volume}{2} (\bibinfo{year}{2016}).
\newblock


\bibitem[\protect\citeauthoryear{Van~Maanen and Marewski}{Van~Maanen and
  Marewski}{2009}]%
        {van2009recommender}
\bibfield{author}{\bibinfo{person}{Leendert Van~Maanen} {and}
  \bibinfo{person}{Julian~N Marewski}.} \bibinfo{year}{2009}\natexlab{}.
\newblock \showarticletitle{Recommender systems for literature selection: A
  competition between decision making and memory models}. In
  \bibinfo{booktitle}{\emph{Proceedings of the 31st Annual Conference of the
  Cognitive Science Society}}. \bibinfo{pages}{2914--2919}.
\newblock


\bibitem[\protect\citeauthoryear{Wheeler}{Wheeler}{2014}]%
        {actrImage}
\bibfield{author}{\bibinfo{person}{Steve Wheeler}.}
  \bibinfo{year}{2014}\natexlab{}.
\newblock \bibinfo{title}{Learning Theories: Adaptive Control of Thought}.
\newblock
\newblock
\newblock
\shownote{[Online under
  \url{http://www.teachthought.com/learning/theory-cognitive-architecture/};
  accessed 19-December-2019].}


\bibitem[\protect\citeauthoryear{Zajonc}{Zajonc}{1968}]%
        {zajonc1968attitudinal}
\bibfield{author}{\bibinfo{person}{Robert~B Zajonc}.}
  \bibinfo{year}{1968}\natexlab{}.
\newblock \showarticletitle{Attitudinal effects of mere exposure.}
\newblock \bibinfo{journal}{\emph{Journal of personality and social
  psychology}} \bibinfo{volume}{9}, \bibinfo{number}{2p2}
  (\bibinfo{year}{1968}), \bibinfo{pages}{1}.
\newblock


\bibitem[\protect\citeauthoryear{Zangerle and Pichl}{Zangerle and
  Pichl}{2018}]%
        {zangerlecontent}
\bibfield{author}{\bibinfo{person}{Eva Zangerle} {and} \bibinfo{person}{Martin
  Pichl}.} \bibinfo{year}{2018}\natexlab{}.
\newblock \showarticletitle{Content-based User Models: Modeling the Many Faces
  of Musical Preference}. In \bibinfo{booktitle}{\emph{ISMIR'18}}.
\newblock


\end{thebibliography}

\end{document}